\documentclass[%
 reprint,
 amsmath,amssymb,
 aps,
floatfix,
]{revtex4-2}
\usepackage{upgreek}
\usepackage{stfloats}
\usepackage{graphicx}
\usepackage{dcolumn}
\usepackage{bm}
\usepackage{marvosym}

\begin{document}


\title{Compact lithium niobate photonic integrated circuits}

\author{Yan Gao\textsuperscript{1}}
\author{Fuchuan Lei\textsuperscript{1}}
\author{Marcello Girardi\textsuperscript{1}}
\author{Zhichao Ye\textsuperscript{1,2}}
\author{Raphaël Van Laer\textsuperscript{1}}
\author{Victor Torres-Company\textsuperscript{1,\Letter}}
\author{Jochen Schröder\textsuperscript{1,\Letter}}

 \affiliation{\textsuperscript{1}Department of Microtechnology and Nanoscience (MC2), Chalmers University of Technology, Göteborg, Sweden\\
 \textsuperscript{2}Qaleido Photonics, Hangzhou, China.\\\textsuperscript{\Letter}jochen.schroeder@chalmers.se, torresv@chalmers.se}




\begin{abstract}
Lithium niobate (LN) is a promising material for future complex photonic-electronic circuits, with wide applications in fields like communications, sensing, quantum optics, and computation. LN took a great stride toward compact photonic integrated circuits (PICs) with the development of partially-etched LN on insulator (LNOI) waveguides. However, integration density is still limited for future high-compact PICs due to the partial edge nature of their waveguides. Here, we demonstrate a fully-etched LN PIC platform which, for the first time, simultaneously achieves ultra-low propagation loss and compact circuit size. The tightly-confined fully-etched LN waveguides with smooth sidewalls allow us to bring the bending radius down to 20 $\upmu$m (corresponds to 1 THz FSR). We have achieved compact high-$Q$ microring resonators with $Q/V$ of 7.1 $\times$ 10$^{4}$ $\upmu$m$^{-3}$, almost one order of magnitude larger than previous demonstrations. The statistical mean propagation losses of our LN waveguides is 8.5 dB/m (corresponds to mean $Q$-factor of 4.9 $\times$ 10$^{6}$) even with a small bending radius of 40 $\upmu$m. Our compact and ultra-low-loss LN platform shows great potential in future miniaturized multifunctional integration systems. As complementary evidence to show the utility of our platform, we demonstrate soliton microcombs with an ultra-high repetition rate of 500 GHz in LN.
\end{abstract}


\maketitle

\noindent

Lithium niobate (LN) is a promising material in integrated photonics, owing to its unique properties that  simultaneously provide electro-optic (EO), nonlinear, acousto-optic effects, broad optical transparency window, coupled with a relatively high refractive index 
 \cite{zhu2021integrated,qi2020integrated,honardoost2020rejuvenating,weis1985lithium}. Applications like low-driving-voltage high-speed EO modulators \cite{wang2018integrated,he2019high}, Kerr, and EO frequency combs \cite{he2019self,zhang2019broadband}, ultra-efficient frequency converters \cite{lu2019periodically,chen2019ultra}, squeezed light sources \cite{nehra2022few}, photon-pair sources \cite{zhao2020high}, and optical parametric oscillators \cite{lu2021ultralow, mckenna2022ultra} have been demonstrated recently.

\begin{figure}[ht] \centering
\centering\includegraphics{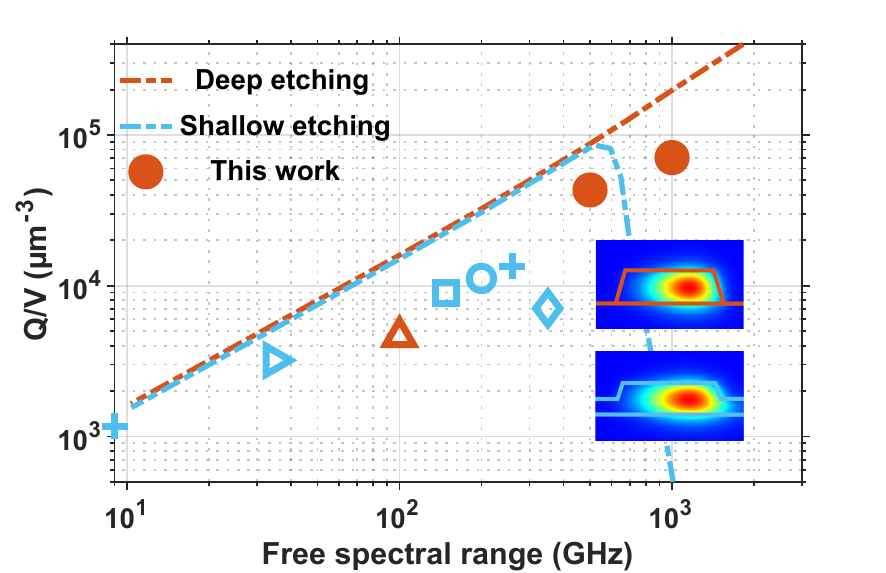}
\caption{Overview of low loss LN integrated platforms. The red color indicates devices with full etching and the blue color indicates devices with shallow etching. The mode volume is calculated via FEM simulation (COMSOL) and the $Q$s are measurement results in the literature. The dashed curve indicates the predicted highest quality over volume ratio $Q/V$ one could achieve for tipical waveguide sizes 2000$\times$600 nm and 2000$\times$300(600) nm. The $Q$s in the prediction are calculated by considering the assumed straight waveguide propagation loss of 2.7 dB/m ($Q$ = 10 million) and simulated bending-caused radiation losses. The references for the comparisons from left to right are: cross \cite{zhang2017monolithic}, right triangle \cite{zhuang2022high}, top triangle \cite{li2022tightly}, square \cite{shams2022reduced}, circle \cite{he2019self}, diamond \cite{gong2020near}. Note the $Q$s used for the comparison corresponds to the statistical mean $Q$s for \cite{li2022tightly}, \cite{shams2022reduced} and this work; while a single-resonance value for the rest.}
\label{fig1}
\end{figure}

For future LN integrated circuits, increased system complexity generally requires large-scale arrays of photonic components \cite{boes2023lithium}. As a result, an ideal integrated LN waveguide platform features both dense integration and low propagation losses. Different LN waveguide configurations have been proposed for a better waveguide platform. Traditionally, LN waveguides are formed in bulk LN based on titanium (Ti) in-diffusion or proton exchange \cite{bazzan2015optical}. The main drawbacks are the large mode area and millimeter-scale bending radius which make dense integration impossible. More recently, the focus has been on shallow etching techniques in thin film LN and great progress has been made on this \cite{zhang2017monolithic}. The state-of-the-art results for shallow-etched rib LN waveguides already exhibit minimum propagation losses at straight sections down to 2.7 dB/m when measuring relatively large ($\sim$ 1.4 cm) race-track resonators \cite{zhang2017monolithic}. However, losses increase to 9.3 dB/m when directly measuring a micro-ring resonator with a bending radius of 80 $\upmu$m \cite{zhang2017monolithic}. The large losses are suspected to come from the increased roughness-caused scattering losses owing to the weak lateral confinement of rib waveguides \cite{zhang2017monolithic} and will prevent further reducing the bending radius for compact circuits.

\begin{figure*}[ht]
\centering\includegraphics[width=0.85\textwidth]{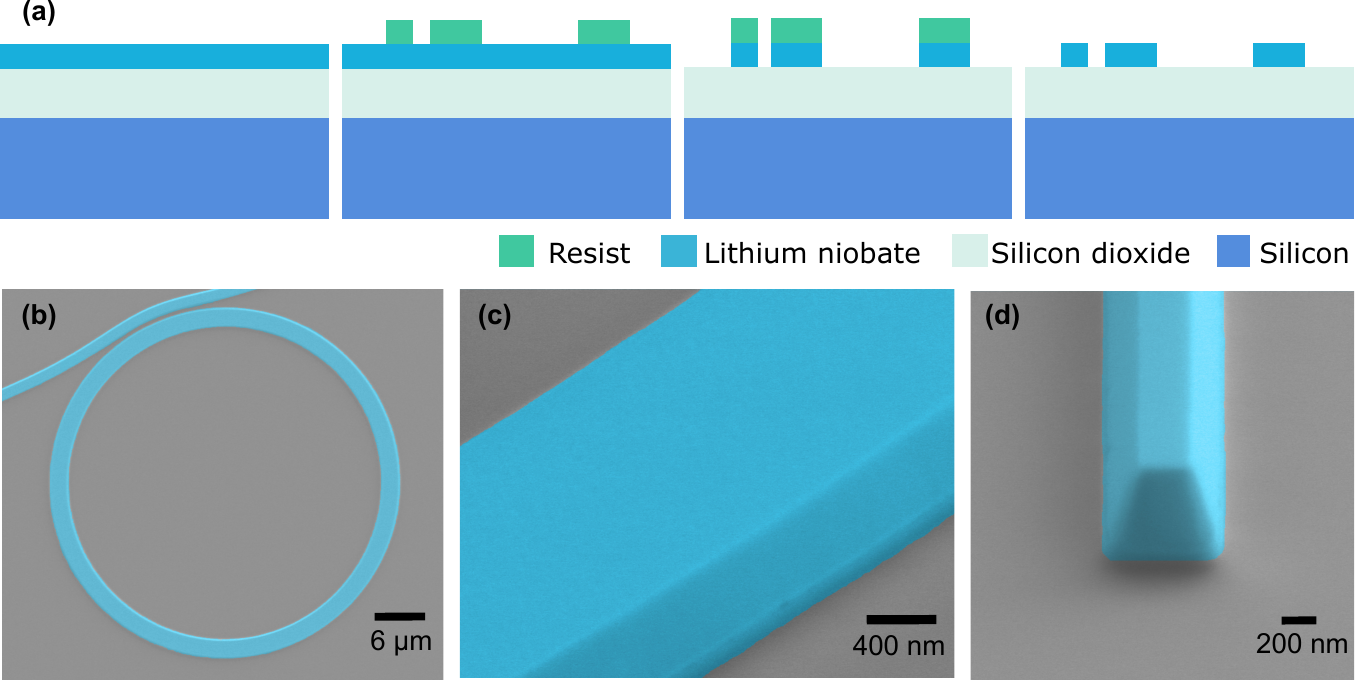}
\caption{(a) Fabrication flow chart for fully etched LN. (b) Top view SEM image of LN ring resonators bending radius of 20 $\upmu$m. (c),(d) High magnification (80 k) colored SEM images of LN waveguides, little roughness can be observed.}
\label{fig2}
\end{figure*}

Towards dense integration, a deeply etched strip waveguide is the best choice since it would significantly improve light confinement both horizontally and vertically thus reducing circuit size compared with shallow-etched rib waveguides. Fully etching could also avoid the uncertainty of etching depth inherent in shallow etching which is beneficial for precise dispersion engineering. However, deeply etched LN waveguides with ultra-low-loss and compact size have until now remained out of reach mainly due to the inherent difficulty of LN deep etching. The difficulty primarily comes from the bad selectivity between LN and its etching masks including both polymer or organic soft resist and dielectric or metallic hard masks 
 \cite{gloersen1975ion,kaufmann2023redeposition}. Deep etching will also raise the interaction between the waveguide mode and rough sidewalls thus requiring increased fabrication quality for a similar propagation loss. Very recently, a diamond-like carbon (DLC) hard mask was used for LN deep etching to improve etching selectivity \cite{li2022tightly}. Low-loss fully-etched LN waveguide has been achieved for a strong multimode waveguide (3300$\times$700 nm), while compact circuits with a few tens micrometers bending radius have not been experimentally demonstrated \cite{li2022tightly}. Investigations \cite{shams2022reduced} show that the material absorption loss for thin film LN is relatively low and the propagation losses for LN waveguides are mainly limited by the sidewall roughness induced scattering losses. Thus, for future compact low-loss LN-integrated platforms, deeply etched strip waveguides with reduced sidewall roughness are still in demand.
 
\begin{figure*}[ht]
\centering\includegraphics[width=0.85\textwidth]{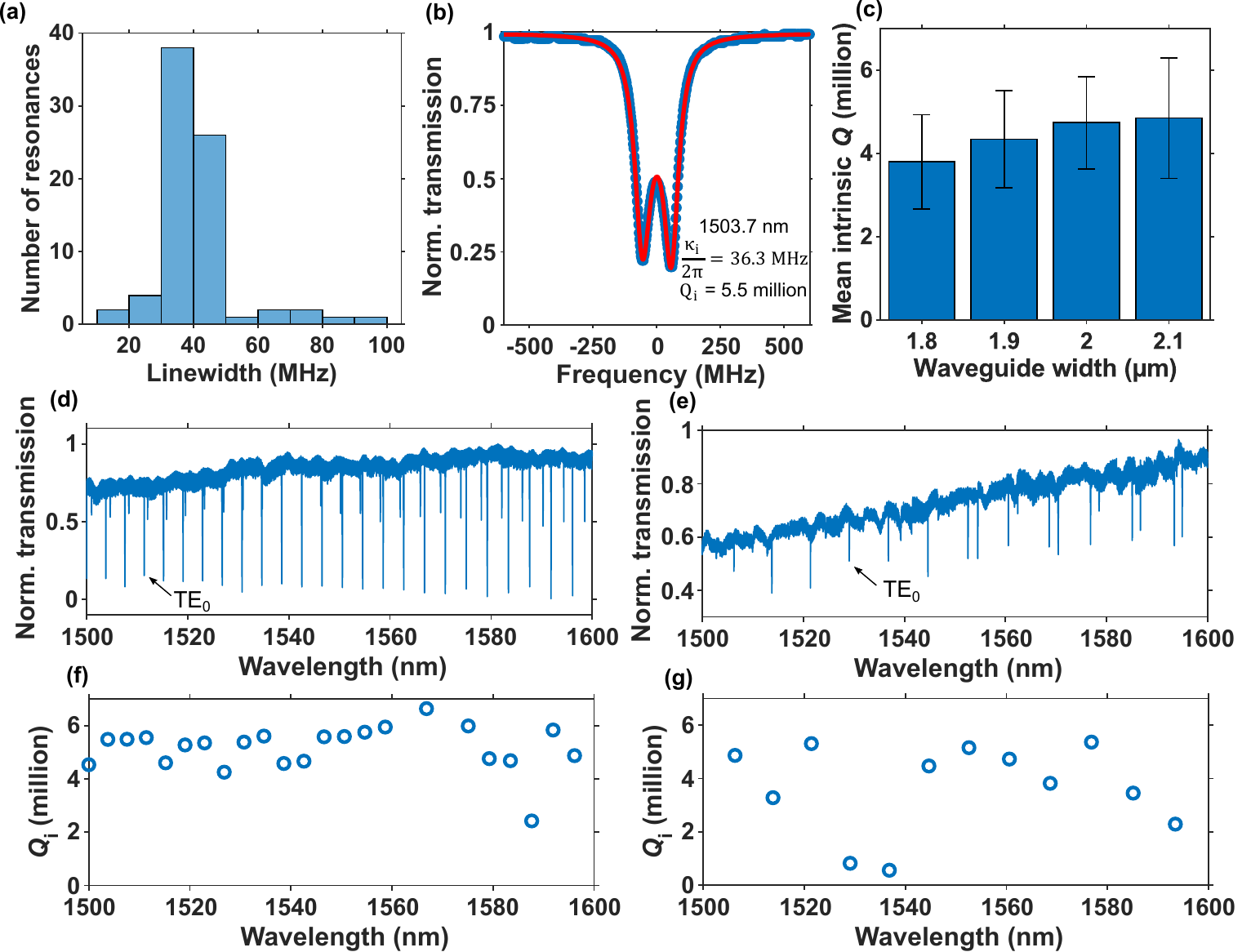}
\caption{(a) Histogram of intrinsic linewidth of resonances from several devices with the same waveguide width of 2.1 $\upmu$m and radius of 40 $\upmu$m. (b) The fitted spectrum of a typical resonance at 1503.7 nm in (d). (c) Statistical intrinsic $Q$s with error bars for devices of waveguide width from 1.8 - 2.1 $\upmu$m and radius of 40 $\upmu$m. (d) A typical transmission spectrum for a ring resonator with 500 GHz (radius 40 $\upmu$m) FSR and (f) its intrinsic $Q$s. (e), (g) Similar spectrum and $Q$s for 1 THz (radius 20 $\upmu$m) ring resonator.}
\label{fig3}
\end{figure*}

Here we show a compact ultra-low-loss deeply etched LN strip waveguide platform, with mean propagation losses as low as 8.5 dB/m (calculated from the statistical intrinsic mean $Q\sim$ 4.9 $\times$ 10$^{6}$ of 500 GHz microrings). Thanks to the smooth waveguide sidewalls and the high light confinement, the waveguide bending radius in LN  could be reduced to 20 $\upmu$m (1 THz FSR) for the first time while maintaining low propagation losses around 11.9 dB/m. Our deeply etched LN platform provides, to the best of our knowledge, the highest $Q/V$ of 7.1 $\times$ 10$^{4}$ $\upmu$m$^{-3}$, almost 1 order of magnitude higher than other results (see Fig. \ref{fig1}.). Here the cavity mode volume \textit{V} is defined: \textit{V} = $\frac{\int\varepsilon\textit{E}^2d\textit{V}}{max(\varepsilon\textit{E}^2)}$. Our high $Q/V$ devices feature excellent light confinement both in temporal and spatial domains. Compared with state-of-the-art results from \cite{zhang2017monolithic}, our compact LN platform achieves a fourfold reduction in bending radius and a 16-fold improvement in integrating density while maintaining the optical losses at the same level. These low-loss compact devices show high potential for future multi-functional and miniaturized optical integrating systems. We utilize these new compact and high-\textit{Q} LN devices to demonstrate to date the highest repetition rate LN solition microcombs.

Starting from Z-cut LN wafers (NANOLN) with LN film thickness of 600 nm and thermally oxidized $\rm{SiO_2}$ of 4.7 $\upmu$m, the sample is prepared via solvent cleaning and standard cleaning (SC1) before spin-coating with MaN 2400 resist. The pattern is first defined on the resist via 100-kV E-Beam exposure (Raith EBPG 5200), in which multipass exposure is used to reduce waveguide sidewall roughness. The sample is then dry-etched in the reactive ion beam etching tool (IBE, Oxford Ionfab 300 Plus) with only Ar plasma. The sidewall angle is estimated around 70 degrees. The etching rate for LN is around 14 nm/min and 45 minutes is required to achieve slight over-etching of 600 nm LN. The resist is then removed via another run of solvent and SC1 cleaning. The SC1 cleaning after etching will also help remove byproducts from ion beam etching, and the slight attack to LN from the cleaning solution may help improve its waveguide sidewall roughness. Unlike some hard-masks-based etching, the LN fabrication flow here only includes one run exposure/etching (see Fig. \ref{fig2} (a)).

A post-colored scanning electron micrograph (SEM) figure of a typical fabricated LN ring resonator with a bending radius of 20 $\upmu$m is shown in Fig. \ref{fig2}(b). Structure defects are hardly seen in SEM figures (Fig. \ref{fig2}(c)-(d)) even at high magnification (80k). The smooth waveguide sidewall obtained significantly reduces scattering losses. Microring resonators with different waveguide widths are fabricated and the $Q$s are characterized to show the high quality of our deeply etched LN waveguides. The ring resonator is coupled with a bus waveguide with a 5° Pully coupler \cite{hosseini2010systematic}. The ring waveguide widths are chosen near 2 $\upmu$m, so that the dispersion is near zero and allows us to obtain both normal and anomalous dispersion waveguides. Laser scanning spectroscopy \cite{del2009frequency,twayana2021frequency} is used to measure the transmission spectrum and the resonance split model \cite{gorodetsky2000rayleigh} is used to fit the resonance to extract the $Q$s and dispersion. The statistical mean $Q$s with error bars for different waveguide widths are shown in Fig. \ref{fig3}(c). The $Q$s of resonances are measured from a few different ring resonators (40 $\upmu$m radius) with the same waveguide width over the wavelength range of 1500-1600 nm. The error bars are calculated based on the standard deviation considering all resonances for statistics. The propagation losses ($Q$s) will decrease (increase) for wider waveguides since the mode will be more confined in the central region and cause less interaction with the sidewall roughness for a wider width. The mean $Q$s of our devices are 3.8-4.9 million with waveguide width from 1.8-2.1 $\upmu$m. For a waveguide width of 2.1 $\upmu$m, the mean $Q$ is 4.9 million, corresponding to a mean propagation loss of only 8.5 dB/m, with a measured group refractive index $n_{g}$ = 2.378. Fig. \ref{fig3}(a) shows the statistical intrinsic linewidth $\upkappa_i$ of the resonances for a waveguide width of 2.1 $\upmu$m and the most probable intrinsic linewidth of 35 MHz is obtained. A typical spectrum for such a device is shown in Fig. \ref{fig3}(d) and the zoomed-in peak with its resonance split fitting around 1503.7 nm is shown in Fig. \ref{fig3}(b), indicating a typical high-$Q$ of 5.5 million. $Q$s over wavelength for the device are shown in Fig. \ref{fig3}(f), from which we can see the $Q$s are high over the 100 nm measurement wavelength range. Similar spectrum and $Q$s for 1 THz (radius 20 $\upmu$m) ring resonators are shown in Fig. \ref{fig3}(e), \ref{fig3}(g). The $Q$s for 20 $\upmu$m microring decrease slightly compared with 40 $\upmu$m one because a small bending radius will push the light mode toward the outside sidewall thus increasing scattering losses.

\begin{figure}[ht]
\centering\includegraphics[width=0.45\textwidth]{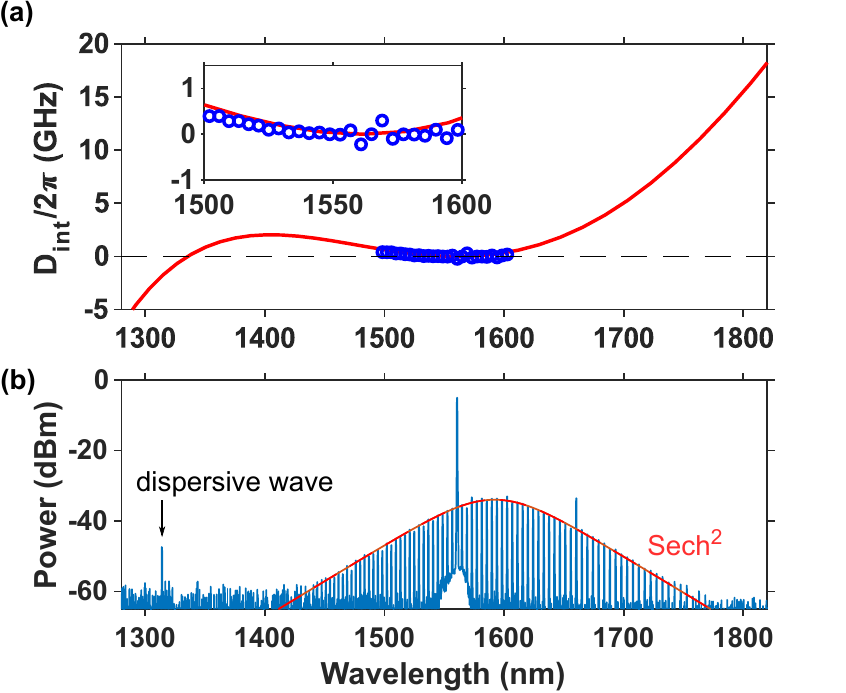}
\caption{(a) Simulated (red) and measured (blue) integrated dispersion of waveguide width of 1.8 $\upmu$m. (b) single soliton spectrum with a repetition rate of ~500 GHz.}
\label{fig4}
\end{figure}

On-chip coherently pumped solitons have attracted significant interest and studied widely in material platforms like silica \cite{yi2015soliton}, silicon \cite{yu2016mode}, and silicon nitride \cite{brasch2016photonic,ye2022integrated}. Recently, soliton microcombs based on LN were also demonstrated and new phenomena like bidirectional switching and self-starting \cite{he2019self, gong2019soliton} were observed. Our low-loss and highly compact LN waveguides are desirable for soliton combs generation and allow to significantly increase the frequency-spacing/repetition rate of the soliton pulse train. To show the potential of our compact LN fully-etched waveguide platform, a 0.5 THz soliton microcomb is demonstrated. We use a 40 $\upmu$m radius microring resonator with a ring waveguide width of 1.8 $\upmu$m for comb generation. The mean $Q$-factor for these devices (shown in Fig. \ref{fig3}(c).) is 3.8 $\times$ 10$^{6}$. Fig. \ref{fig4}(a) shows the simulated and measured integrated dispersion $D_{\rm{int}} = \omega_\mu - \omega_0 - \mu D_1 $, and the extracted group velocity dispersion $\beta_2$ is -10 $\rm{ps^2/km} $. The device is designed to achieve small anomalous dispersion in order to obtain a bright soliton state \cite{herr2014temporal}. A Toptica laser (CTL 1550) amplified by EDFA is used to pump the ring resonator at a wavelength of 1560.7 nm. By manually tuning the laser across the resonance, a broadband soliton microcomb is obtained without requiring fast tuning technology or external triggering, which are attributed to the photorefractive effect in LN \cite{he2019self, gong2020near}. A single-soliton comb is measured via an optical spectrum analyzer and the result is shown in Fig. \ref{fig4} (b). In Fig. \ref{fig4} (b), a smooth sech$^2$ shape envelop of the comb spectrum is obtained, and a dispersive wave is also observed near 1300 nm which could be predicted from simulated dispersion in Fig. \ref{fig4} (a).

In conclusion, a fully-etched strip LN waveguide platform for integrated photonics has been demonstrated with advantages for simultaneously achieving ultra-low propagation losses and ultra-small circuit size. The fabrication process only includes one run exposure/etching with a commonly used e-beam resist. Our high-quality LN waveguides with smooth sidewall roughness and tight-confining deeply etched structure allow us to achieve unprecedented compact devices with ultra-low propagation losses. A 16-fold improvement in integration density has been experimentally demonstrated compared with state-of-the-art results with the same low-loss level. As a result, terahertz high-$Q$ microring resonators with a bending radius of 20 $\upmu$m are possible and have been demonstrated for the first time in LN. Our ring resonators show almost one order of magnitude improvement of $Q/V$ compared with other results. The high-$Q/V$ microring resonators also enable broadband soliton microcombs generation with the highest repetition rate of 0.5 THz in LN. The low-loss LN waveguide platform could be used in most applications situations like E-O devices, nonlinear frequency conversion, and communications. Combined with its high compactness, the platform is especially desirable for situations where large-amount components are required to achieve their complex functionality such as large-scale optical switching arrays \cite{seok2019wafer}, integrated photonic LiDAR \cite{zhang2022large}, and optical artificial intelligence \cite{zhang2021optical}. Nevertheless, our high $Q/V$ optical resonators with excellent temporal and spatial light confinements will also be beneficial for efficient nonlinear processes,  large finesse optical bio-chemical sensors, and cavity quantum electrodynamics (cQED) studies \cite{vahala2003optical}. In the future, to achieve even lower-loss compact LN PICs, the fabrication flow should be further optimized to reduce sidewall roughness.

\vspace{0.4cm}
\noindent
$\mathbf{Funding:}$ Swedish Research council (VR-2017-05157, VR-2021-04241, VR-2016-06077);

\vspace{0.2cm}
\noindent
 $\mathbf{Acknowledgments:}$ The authors thank cleanroom staff from Myfab at Chalmers Nanofabrication Laboratory for discussion and training.

\bibliography{sample}



\end{document}